\documentstyle[conf_iap]{article}
\def\hmpc{\;h^{-1}{\rm Mpc}}
\def\hgpc{\;h^{-1}{\rm Gpc}}
\def\lya{Ly$\alpha$}
\begin{document}

\heading{%
Recovery of the mass power spectrum from the Ly-$\alpha$ 
forest: The effect of ionizing background fluctuations from discrete QSO
sources.
%
} 
\par\medskip\noindent
\author{%
Rupert A.C. Croft
}
\noindent{\it
Dept. of Astronomy, Ohio State University, Columbus, OH 43210,
USA.}

\begin{abstract}
In gravitational instability
 models,  there is a close relationship between 
most  \lya\  absorption and the matter density along the line of sight.
Croft et al. (1997) have shown that it is possible to use this
 relationship with
some additional assumptions (such as Gaussianity of the initial density) to
recover the power spectrum of mass fluctuations, $P(k)$, from QSO absorption 
spectra. The relative uniformity of the ionizing radiation background on the
scales of interest is  an additional assumption required by the technique.
Here we examine the fluctuations in \lya\ spectra 
caused if the ionizing background  radiation is generated by discrete QSO
sources. We present our results alongside the preliminary application of
the $P(k)$ recovery technique to the QSO Q1422+231. 
We find the ionizing background fluctuations to have an effect orders of
magnitude smaller than the matter fluctuations at the scales on which we
measure $P(k)$ ($\lambda < 10 \hmpc$).

\end{abstract}

\vspace{0.2cm}

In the ``Fluctuating Gunn-Peterson Approximation'' for the high-$z$ \lya\
forest (see the contribution by Weinberg in these proceedings), variations in
the optical depth along the line of sight to a QSO are directly
related to fluctuations both in the underlying matter density and in the
intensity of the ionizing background radiation. Croft et al. (1997)
 (\cite{CWKH}) have developed a
technique to recover $P(k)$ for the mass, which assumes that only the matter
fluctuations have a non-negligible effect. The results of an illustrative
application to a Keck spectrum of QSO Q1422+231 (observation by
\cite{Son}) are shown in Figure 1(a).

 In order to test at least qualitatively 
the effect of ionizing background flucuations, which were not included in
\cite{CWKH} , we have created simulated QSO \lya\ absorption spectra
in a universe where the matter density is uniform and the only fluctuations
present in absorption are those due to the sources of ionizing radiation
 being 
discrete. We have populated a $0.6 \hgpc$ comoving cube (in a
 $q_{0}=0.1$, h$_{100}$=0.5 universe) with QSOs distributed
randomly and with luminosities randomly sampled from the ($z=3$) luminosity
function of \cite{HM}. We include a lower cut-off at $M_{B}=-23$
and calculate the opacity of the IGM in Euclidean space in the 
optically thin limit. Five randomly chosen example spectra (each is $\sim0.5$
times the length of the cube side) are shown in Figure
1(b). The mean absorbed flux, $D_{A}$, for all spectra
was chosen to be the same as $D_{A}$ for Q1422+231, which is also shown in
Figure 1(b). The line marked ``$\lambda_{max}$'' marks
 the largest scale at which
we have measured $P(k)$ in Figure 1(a). We can see that fluctuations due to the
background are indeed small on this scale. We find that the power
spectrum of the transmitted flux (F) on this scale in
 the models is $2\%$ of the 
value for Q1422+231. For the next largest of the k-modes plotted, the ratio is
$0.3 \%$ .
  
This simulation of the effects of discrete QSO sources is 
extremely simplified, and does not include the effects of redshift (except
for the finite box size, which cuts off QSO flux beyond $\Delta z \sim 0.8$),
 QSO clustering or optically thick absorption. However, we expect that  any 
additional radiation fluctuations due to these effects will not change
our conclusion that the matter fluctuations are dominant enough
(being of order unity) on these
scales that we can infer $P(k)$ for the mass reliably.
If there is a contribution to the ionizing background from other sources with
greater number density than QSOs, such as starburst galaxies or Population III
stars, we expect the 
radiation fluctuations to be even smaller. If the source positions
are correlated with the absorbing gas in this case, these smaller
fluctuations could have more of an effect on the $P(k)$ measurement, although
as the relative range of the ``proximity effect'' due to these objects will be
so small, we expect the effect to still be unimportant.

\begin{figure}
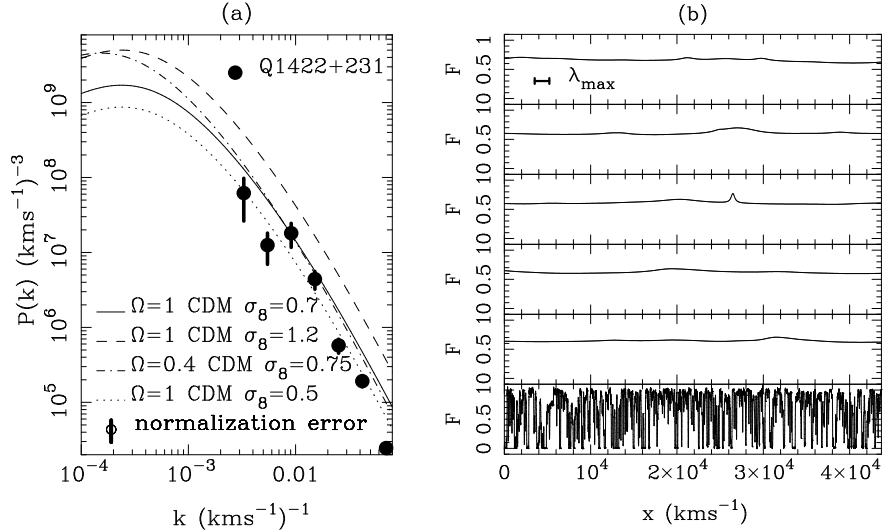

\centering
\vspace{6.0cm}
\includegraphics{pkqso.ps}
\includegraphics{gpspec.ps}
\caption[]{
(a) The mass power spectrum recovered from the \lya\ forest
of QSO Q1422+231 (points). The lines show the predictions of linear theory
for 4 different CDM models.
(b) The top 5 panels show randomly chosen lines of sight representing
QSO absorption spectra taken through a universe where the IGM is homogenous 
and uniform in density and where the ionizing background is generated by
discrete QSO sources. The bottom panel shows the absorption spectrum of
Q1422+231 on the same scale. The line marked ``$\lambda_{max}$'' is the
wavelength corresponding to the largest scale point plotted in Figure 1(a).
}
\end{figure}

\acknowledgements{ }
I thank A. Songaila and L. Cowie for providing the spectrum of Q1422+231
and D. Weinberg, L. Hernquist and N. Katz for permission to discuss our 
joint work.

\vfill
\end{document}